\documentclass[aps,prd,twocolumn,groupedaddress,showpacs,nofootinbib,amssymb,balancelastpage,preprintnumbers,floatfix]{revtex4}

\usepackage[dvipdfmx]{graphicx}
\usepackage{graphicx,bm,color}

\usepackage{amsmath}
\usepackage{amssymb}
\usepackage{amsfonts}
\usepackage{cases}
\usepackage{cancel}
\usepackage{hyperref}
\usepackage[normalem]{ulem}
\usepackage{subfigure}
\usepackage{here}
\usepackage{color}
\usepackage{comment}

\usepackage{tikz}
\usetikzlibrary{matrix}

\allowdisplaybreaks[3]

\newcommand{\nLk}{n_{L, \vec{k}}^s}
\newcommand{\nRk}{n_{R, \vec{k}}^s}
\newcommand{\DLk}{D_{L, \vec{k}}^s}
\newcommand{\DRk}{D_{R, \vec{k}}^s}
\newcommand{\nLmk}{n_{L, -\vec{k}}^s}
\newcommand{\nRmk}{n_{R, -\vec{k}}^s}
\newcommand{\DLmk}{D_{L, -\vec{k}}^s}
\newcommand{\DRmk}{D_{R, -\vec{k}}^s}

\newcommand{\nLkd}{n_{L, \vec{k}}^{s \dagger}}
\newcommand{\nRkd}{n_{R, \vec{k}}^{s \dagger}}
\newcommand{\DLkd}{D_{L, \vec{k}}^{s \dagger}}
\newcommand{\DRkd}{D_{R, \vec{k}}^{s \dagger}}
\newcommand{\nLmkd}{n_{L, -\vec{k}}^{s \dagger}}
\newcommand{\nRmkd}{n_{R, -\vec{k}}^{s \dagger}}
\newcommand{\DLmkd}{D_{L, -\vec{k}}^{s \dagger}}
\newcommand{\DRmkd}{D_{R, -\vec{k}}^{s \dagger}}
\newcommand{\eiqk}{e^{i \theta_{\vec{k}}}}
\newcommand{\eiqmk}{e^{-i \theta_{\vec{k}}}}
\newcommand{\abk}{\left| \vec{k} \right|}
\newcommand{\tilmN}{\tilde{m}_N}
\newcommand{\tilmNs}{\tilde{m}_N^\ast}
\newcommand{\mD}{m_D}
\newcommand{\mDs}{m_D^\ast}
\newcommand{\MR}{M_R}
\newcommand{\MRs}{M_R^\ast}
\newcommand{\ML}{M_L}
\newcommand{\MLs}{M_L^\ast}

\newcommand{\pt}{\partial_t}

\begin{document}

\title{
QCD preheating: New frontier of baryogenesis
}

\author{Xin-Ru Wang}\thanks{{\tt  wxr21@mails.jlu.edu.cn}}
\affiliation{Center for Theoretical Physics and College of Physics, Jilin University, Changchun, 130012,
China}

\author{Jin-Yang Li}\thanks{{\tt lijy1118@mails.jlu.edu.cn}}
\affiliation{Center for Theoretical Physics and College of Physics, Jilin University, Changchun, 130012,
China} 
\affiliation{KEK Theory Center, IPNS, Tsukuba, Ibaraki 305-0801, Japan}

\author{Seishi Enomoto}\thanks{{\tt seishi@mail.sysu.edu.cn}}
	\affiliation{School of Physics, Sun Yat-Sen University, Guangzhou 510275, China}

\author{Hiroyuki Ishida}\thanks{{\tt ishidah@pu-toyama.ac.jp}}
	\affiliation{Center for Liberal Arts and Sciences, Toyama Prefectural University, Toyama 939-0398, Japan}

\author{Shinya Matsuzaki}\thanks{{\tt synya@jlu.edu.cn}}
\affiliation{Center for Theoretical Physics and College of Physics, Jilin University, Changchun, 130012,
China}

\begin{abstract}  

We find that 
QCD can create the cosmological matter abundance via out-of-equilibrium processes during the QCD phase transition, that is what we call the QCD preheating, 
where the dynamic transition of the QCD vacuum characterized by the quark condensate takes place instantaneously.  
\textcolor{black}{
This mechanism works when the Universe undergoes subsequent supercooled QCD transition.} 
We also find that the QCD preheating can work to create the baryon asymmetry of the Universe if there is the new physics communicated with QCD. 
These are new pictures of the thermal history around the QCD-phase transition epoch, and thus the dynamic aspect of the QCD vacuum opens a new frontier to explore low-scale matter generation such as baryogenesis. 
Pursuing the QCD reheating era would also help deeply understanding the subatomic-scale physics in the thermal history of the Universe. 

\end{abstract} 
\maketitle


%
\section{Introduction}

The QCD phase transition is of importance to understand the origin of mass and property for matter, i.e., nucleon, and has extensively been explored. 
\textcolor{black}{
In application to cosmology, however, 
QCD does not play major role to account for cosmological matter abundances observed in the Universe today including the baryon asymmetry of the Universe. 
This is simply because 
the QCD sector does not have baryon number violation nor large enough CP violation, and will never be out of thermal equilibrium at lower scales due to the strong coupling nature.  
This is ``folklore" that is widely accepted, and thus, 
people tend to call for Beyond the Standard Model, which is usually setup irrespective to the QCD phase transition~\footnote{
\textcolor{black}{QCD baryogenesis has been argued in~\cite{Ipek:2018lhm,Servant:2014bla,Croon:2019ugf} with higher scale confinement, not at the ordinary QCD scale.}}. 
}

In this paper, 
\textcolor{black}{we claim that} 
{\it dynamic} aspects of the QCD vacuum during the 
QCD phase transition can be crucial to 
create cosmological matter abundances observed in the Universe today: the time-varying light-quark condensate 
$\langle \bar{q}q \rangle $ produces the total matter-antimatter abundance, through undergoing a supercooling and reheating 
before the static hadron phase is formed. 

The dynamic aspect of vacuum in quantum field theory 
has extensively been explored so far in light of the reheating process
after the inflation in the early Universe, 
so-called preheating~\cite{Dolgov:1989us,Traschen:1990sw,Kofman:1994rk,Shtanov:1994ce,Kofman:1997yn}. 
(for reviews, see e.g., \cite{Kofman:1997yn,Amin:2014eta,Lozanov:2019jxc,Enomoto:2020aapps}).
The oscillating background field of the inflaton induces the nonadiabatic state for other coupled species. 
It drives the nonperturbative particle production,  and the amount is 
exponentially amplified by the parametric resonance similar to a swing with a pumping oscillator.

The ``ballpark" of preheating has also been extended to the  
baryogenesis via the nonperturbative production at high scales \cite{Funakubo:2000us,Rangarajan:2001yu,Pearce:2015nga,Enomoto:2017rvc,Wu:2019ohx,Enomoto:2020lpf,Lee:2020yaj}. 
The mechanism of preheating is applied to interesting variant scenarios:
the production of massive particles heavier than the inflaton \cite{Kolb:1996jt,Kolb:1998he,Giudice:1999fb,Peloso:2000hy}; 
the preheating due to the alternative field instead of the inflaton
\cite{Garcia-Bellido:2008ycs,Adshead:2015kza,Ema:2016dny,Adshead:2018oaa}, etc.

Preheating at the subatomic scale has never 
been discussed in the context of baryogenesis. 
A scalar condensate is present even in QCD, that is 
the light quark condensate $\langle \bar{q}q \rangle $, 
and it can couple to the nucleon state as well as meson states 
in a systematic way respecting the chiral symmetry and its breaking 
by the current quark masses. 
Therefore, 
the dynamic motion of $\langle \bar{q}q \rangle $ should have 
the potential to explosively produce the number densities for nucleon and anti-nucleon by nonadiabatic processes, 
similarly to the preheating induced by the nonadiabatic-varying vacuum. 
In this paper, we call the observation above the {\it QCD preheating} and find it indeed work.

Remarkably enough, the QCD preheating reveals a new picture of the thermal history related to nucleon: 
cosmological matters are produced out of equilibrium and overwhelm the thermal abundance, which survives long enough during the reheating epoch. 
This new picture of the thermal history is applicable to the baryogenesis at the subatomic scale once the QCD preheating is coupled to Beyond the Standard Model, that is the {\it QCD-scale baryogenesis}.

\section{Overview of the QCD preheating scenario} 
\begin{figure}[t]
    \centering
     \includegraphics[scale=0.35]{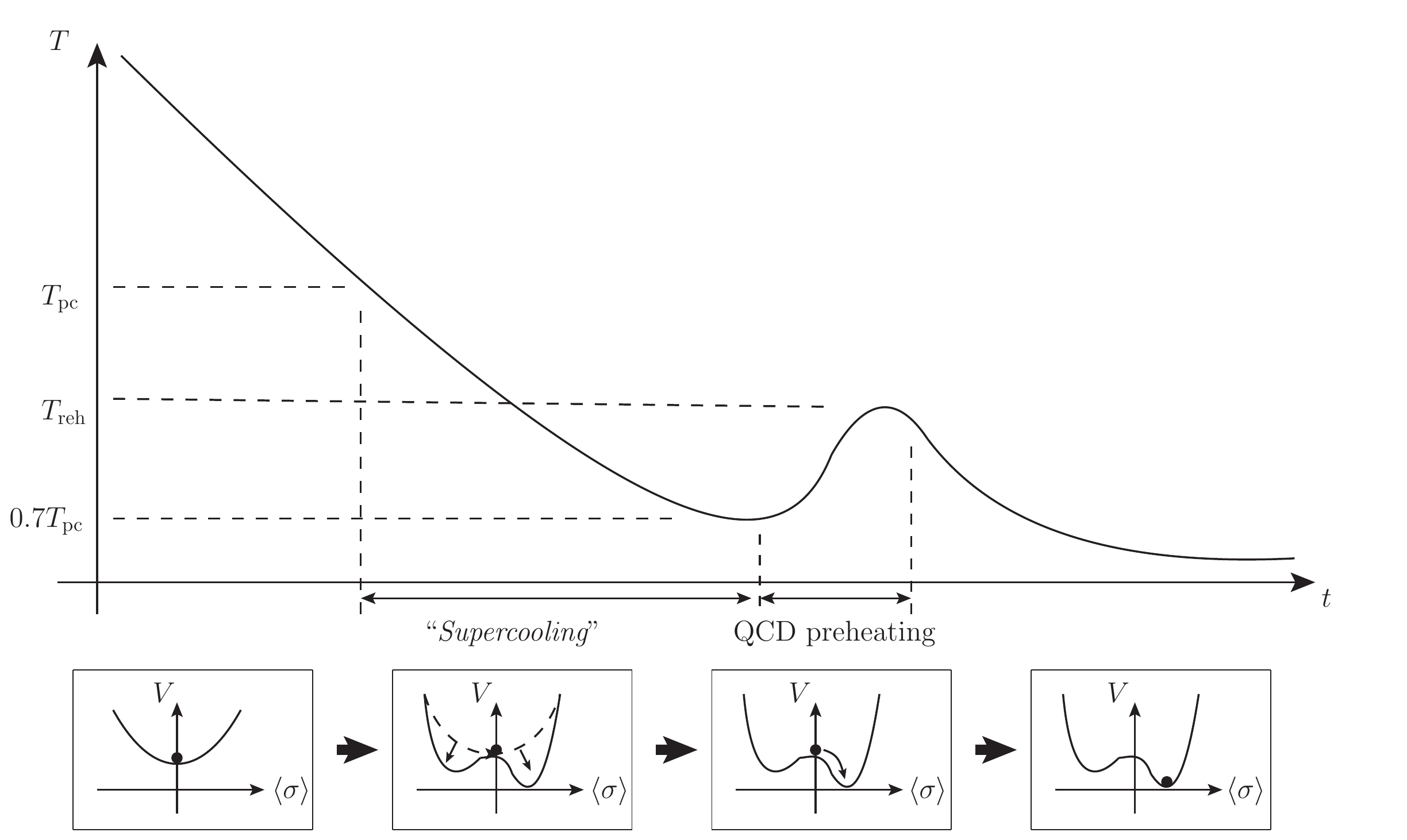}
    \caption{A schematic view of the QCD preheating 
    in the thermal history of the Universe. The top panel depicts the evolution of the temperature of the Universe at around the QCD phase transition, where the QCD preheating via the dynamic $\langle \sigma \rangle  \sim \langle \bar{q}q \rangle $ happens 
    at, say, $T=T_n \sim 0.7 T_{\rm pc}\simeq 109$ MeV by starting the roll-down in the potential ($V$)  passing the supercooling, and reheats the Universe 
     up to $T = T_{\rm reh} \simeq 119$ MeV 
    (see the text). 
    \textcolor{black}{
    The bottom panel traces the time evolution of $T$ in terms of the potential shapes of the dynamic $\langle \bar{q}q \rangle$ observed in the reference frame comoving with the Hubble parameter. 
    Since the $\langle \sigma \rangle$ motion is much faster than the Hubble expansion rate,  
    a sudden discontinuous jump-up of $\langle \sigma \rangle $, at $T=T_n$ 
would be detected by eyes of observers getting on the Hubble reference frame,  
which can be interpreted as a first order transition in the thermal history of the Universe. 
}
Passing the relaxation as in Eq.(\ref{relax:time}), the system 
    comes back to the normal thermal equilibrium, which makes the potential of $\langle \sigma \rangle $ lifted up in the standard manner of thermal QCD.  
    }
    \label{rough_sketch}
\end{figure}
To begin with, we outline the static aspect of the QCD chiral phase transition, and 
\textcolor{black}{
turn to a possible dynamic feature of the QCD vacuum}.

The lattice QCD has confirmed that thermal QCD with 2 + 1 flavors at the physical point undergoes the crossover for 
the chiral phase transition, what is called the chiral crossover, 
at the pseudocritical temperature $T_{\rm pc}\sim 155$ MeV~\cite{Aoki:2009sc,Borsanyi:2011bn,Ding:2015ona,Bazavov:2018mes,Ding:2020rtq}. 
At almost the same temperature, the deconfinement-confinement transition (crossover) is expected to happen as well~\cite{Bazavov:2016uvm,Ding:2017giu}. 
Above $T_{\rm pc}$, the light quark condensate $\langle \bar{q}q \rangle$ takes nearly vanishing values when properly renormalized 
to be divergent free.  
Cooling down below $T_{\rm pc}$, 
$\langle \bar{q}q \rangle$ starts to get sizable and finally reaches the value measured at the vacuum when $T$ goes well below $T_{\rm pc}$, $T/T_{\rm pc} \lesssim 0.7$~\cite{Aoki:2009sc}.


\textcolor{black}{
This is the static picture of the QCD phase transition at our best knowledge.  
When Hubble evolutionary 
Universe and some possible Beyond the Standard Model contributions 
are taken into account, the dynamic picture of the QCD phase transition 
might drastically be altered. 
In particular, 
subsequent supercooling including the QCD phase transition 
can be realized when the Standard Model is extended to be 
scale invariant along with a dark sector, as discussed in~\cite{Pisarski:1983ms,Iso:2017uuu,Hambye:2018qjv,Sagunski:2023ynd}
~\footnote{
\textcolor{black}{
In the literature~\cite{Kim:2018byy,Kim:2018knr,Kim:2019lks}, it was argued that even without Beyond the Standard Model, 
a barrier-less boundary for a single bubble between the chiral symmetric or deconfining and broken or confining phases is interpreted as the crossover,  which starts to happen at $T=T_{\rm pc}$, that is consistent with what the lattice simulations have observed. The confining hadron-phase bubbles expand until all bubbles occupy the Universe, 
at which time, the so-called the nucleation and percolation, $\langle \bar{q}q  \rangle $ reaches the true vacuum (at $T \lesssim 0.7 T_{\rm pc}$). 
This implies that the evolution of the hadron phase acts as if the Universe undergoes a {\it supercooling} in the regime $0.7 T_{\rm pc} \lesssim T \lesssim T_{\rm pc}$. 
The temperature where the hadron phase bubbles fully cover the Universe,  $T_f$, 
has been estimated to be indeed lower than 
$T_{\rm pc}$: $T_f/T_{\rm pc} \sim 0.76$. 
However, in the literature~\cite{Kim:2018byy,Kim:2018knr,Kim:2019lks} the used pressure expressions are not widely accepted, hence do not match with the lattice QCD thermodynamics. 
}}}.

\textcolor{black}{In the literature~\cite{Sagunski:2023ynd} 
the QCD supercooling has been realized by coupling to 
a dark sector which undergoes an ultrasupercooling, 
analogously to the scenario of the QCD-induced supercooling electroweak phase transition~\cite{Iso:2017uuu}. 
This subsequent supercooling is triggered by the highly dominated potential energy density of the dark sector in the Hubble parameter, which can be operative until $T \sim T_{\rm pc}$.  
All the massless six quark condensates are trapped by the thermal potential at the origin (the chiral symmetric phase). 
After the supercooling ends in the dark and the Higgs sectors (via tunneling into the true vacua), 
quark mass terms are generated as usual, and only lightest flavors (say, up and down quarks) survive in 
the QCD thermal plasma. 
Thus the QCD sector also exits from the supercooling.  
}

\textcolor{black}
{Then the light-quark condensate $\sigma \sim \langle \bar{q}q \rangle$ starts to roll down to the true vacuum, 
along the generated linear (negative) slope $\sim -m_q \sigma$ at the origin. 
This rolling motion is much faster than the Hubble rate, so can be nonadiabatic and out-of-equilibrium to 
provide the chance of particle production in the QCD phase transition epoch.
This is the setup that we are concerned about in the present paper. 
}

\textcolor{black}{ 
Realization of such subsequent supercooling involves a nonperturbative analysis 
on QCD with massless six flavors. 
In the literature~\cite{Sagunski:2023ynd},
the authors have worked on chiral effective models in the mean field approximation (only including 
one-loop corrections from quarks), which corresponds to the limit of large number of colors (so-called the large $N_c$ limit) 
and assumed that the usual lightest three flavors predominantly form the quark condensates faster than other quarks.   
Then the nucleation temperature $T_n$ for the completion of the QCD phase transition over the Hubble evolution  
has been estimated 
as $T_n/T_c \sim 0.98$ for a phenomenologically viable 
dark sector setup, where $T_c$ denotes the 
critical temperature for the first order chiral phase 
transition in QCD with massless three flavors. 
This value could be corrected by amount of ${\cal O}(1/N_c)\sim 30\%$ when the subleading contributions 
in the large $N_c$ expansion could be incorporated, which is to be $T_n/T_c \sim 0.7$. 
Furthermore, a larger number of flavors could make the strength of the first order phase transition stronger, which 
would imply the smaller $T_n/T_c$. 
Thus we may suspect that the QCD supercooling could last long enough and $T_n$ is low enough that 
the thermal correction to $\langle \bar{q}q \rangle$ can safely be neglected when it starts to roll down. 
}



\textcolor{black}{
In this paper, we discuss 
the dynamic $\langle \bar{q}q\rangle$ 
by assuming a possible supercooling-like picture  
during the QCD phase transition epoch as above driven by coupling to 
the Beyond the Standard Model. 
The dynamic motion of $\langle \bar{q}q \rangle$ 
instantaneously starts 
at say, $T \sim 0.7 T_{\rm pc}$ passing the supercooling. 
Since the scalar field $\langle \bar{q}q \rangle$ couples to 
the meson and baryon states, the nonadiabatic particle production will then 
take place and reheat the Universe from $0.7 T_{\rm pc}$ up to 
$T_{\rm reh} (< T_{\rm pc})$. 
This is how the QCD preheating works and can successfully 
create various matters and antimatters through the nonperturbative and nonadiabatic processes, as will be explicitly seen below. 
} 

A schematic view of the QCD preheating is illustrated in Fig.~\ref{rough_sketch}.

\section{Dynamic QCD vacuum via  
a chiral effective theory description} 

\textcolor{black}{
To describe the dynamic $\langle \bar{q}q \rangle $ after the QCD supercooling ends as elaborated in Sec.II, 
we employ a chiral effective theory, called the linear sigma model, with the lightest two flavors}, 
in which color-singlet composite 
operators $\bar{q}q$ and $qqq$ are 
monitored by hadronic-interpolating 
fields. 
The model is built based on the chiral $SU(2)_L \times SU(2)_R$ symmetry and its breaking structure with the current quark mass for the up and down quarks included. 
The building blocks are: i) the two-by-two complex scalar matrix: 
$M \sim \bar{q}_R q_L$, which transforms 
under the chiral symmetry as $M \to U_L \cdot M \cdot U_R^\dag$ with  
$ U_{L,R} \in SU(2)_{L,R}$, 
where $M$ is parametrized by the isosinglet sigma ($\sigma$) mode 
and isotriplet pion ($\pi$) mode as 
$M = \sigma\cdot 1_{\bf 2 \times 2}/2 + i \pi^a \tau^a/2$ with the Pauli matrices $\tau^a$ ($a= 1,2,3$); ii) the nucleon (proton, neutron)-doublet field $N_{L,R} = (p, n)^T_{L,R}$, which belong to the fundamental representation of $SU(2)_{L,R}$ groups. 
The linear-sigma model Lagrangian is thus given as 
\begin{align} 
& \mathcal{L} 
= \operatorname{tr}\left[\partial_{\mu} M^{\dagger} \partial^{\mu} M\right] 
- V 
\notag\\ 
& \hspace{20pt} 
+\bar{N} i \partial N-\frac{2 m_{N}}{f_{\pi}}\left(\bar{N}_{L} M N_{R}+\bar{N}_{R} M^{\dagger} N_{L}\right) 
\,, \notag\\ 
& V  = 
m_{\pi}^{2} f_{\pi} \operatorname{tr}\left[{\rm Re}(M) \right] 
+ m^{2} \operatorname{tr}\left[M^{\dagger} M\right] 
+ \lambda\left(\operatorname{tr}\left[M^{\dagger} M\right]\right)^{2}
\,. \label{LSM-Lagrangian:eq}
\end{align}
$f_\pi$ is the pion decay constant $\simeq 92.4$ MeV, 
$m_\pi$ the pion mass $\simeq 140$ MeV,
and $m_N$ the nucleon mass $\simeq 940$ MeV. The vacuum expectation value of $\sigma$, $\langle \sigma \rangle $, and its dynamical evolution are identical to those for $\langle \bar{q}q \rangle $. 
We fix the potential parameters $m^2$ and $\lambda$ to the values determined 
at the true vacuum $\langle \sigma \rangle  = f_\pi$ satisfying the stationary 
condition. Then we have $\lambda f_\pi^2 = (M_\sigma^2 - m_\pi^2)/2 $ 
and $m^2= 3 m_\pi^2/2 - M_\sigma^2/2$, with the $\sigma$ mass squared defined as 
$M_\sigma^2 = \partial^2 V/\partial \sigma^2 |_{\sigma = f_\pi}$, which we take  $\simeq(500 \: {\rm MeV})^2$.

The equation of motion for $\langle \sigma \rangle  \sim \langle \bar{q}q \rangle $  
with the space homogeneity leads  
\begin{equation}
0=\langle\ddot{\sigma}\rangle  +\gamma \langle\dot{\sigma}\rangle  -m_{\pi}^{2} f_{\pi}+m^{2}\langle\sigma\rangle +\lambda\langle\sigma\rangle ^{3} + \cdots 
\,. \label{EOM-sigma} 
\end{equation}
The ellipses denote the negligible terms including the Hubble friction term $(3 H \langle \dot{\sigma} \rangle )$ and the backreactions from the pion and nucleon fields.
$\gamma$ plays the role of the full width of the $\sigma$ meson identified as $f_0(500)$ in the Particle Data Group. 
As a phenomenological input, we take $\gamma$ to be the central value of the current measurement, $\gamma \equiv |2 {\rm Im}[\sqrt{s}_{\rm pole}]| = 550$ MeV~\cite{ParticleDataGroup:2020ssz}.

The dynamic oscillation of $\langle \bar{q}q \rangle$ 
will globally be processed 
\textcolor{black}{ 
along the potential, which is well approximated to be $V$ in Eq.(\ref{LSM-Lagrangian:eq}) 
after the supercooling ends at, say, $T_n = 0.7 T_{\rm pc}$. 
Since the oscillator $\langle \bar{q}q \rangle$ couples to the meson and baryon states, the nonadiabatic particle production will then take place and reheat the universe from $T=T_n$ up to 
the reheating temperature $T_{\rm reh} (< T_{\rm pc})$. 
The typical time scale (1/500, {\rm MeV}$^{-1}$) of the dynamic $\langle \bar{q}q \rangle$  oscillation is much faster than the background Hubble evolution ($t\sim 1/H\sim 1/(T^2/M_p)$ with $T\sim0.1\,{\rm GeV}$ and $M_p \sim10^{19}\,{\rm GeV}$ being the Planck scale). 
Hence the Hubble Universe will observe as if 
$\langle \bar{q}q \rangle$ instantaneously jumps to the true vacuum (hadron phase) from the origin (symmetric phase). 
 }

\begin{figure}[t]
\centering
\includegraphics[scale=0.4]{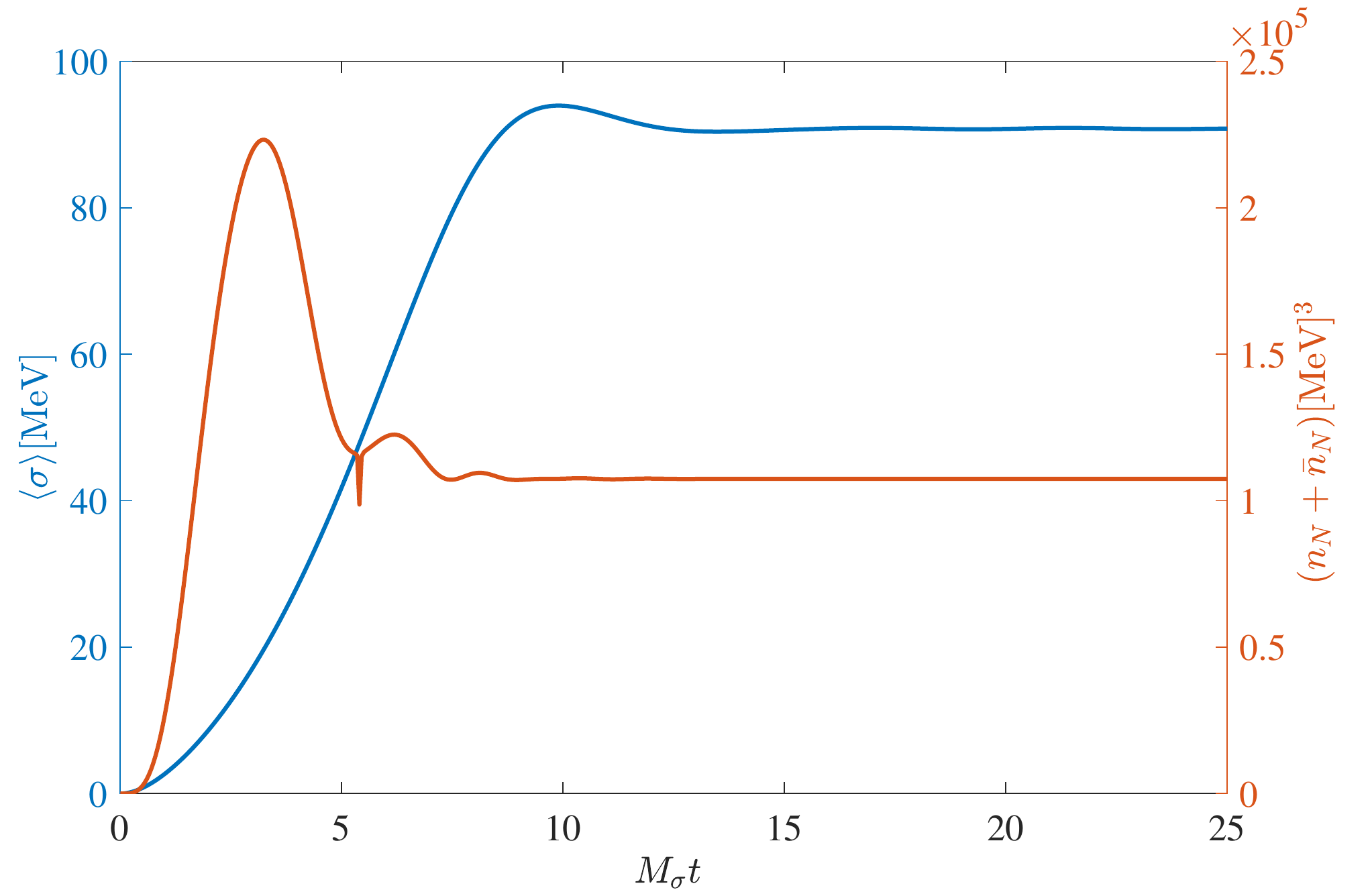}
\caption{The time evolution of $\langle \sigma \rangle $ (blue) and the total nucleon number density $(n_N + \bar{n}_N)$ (red), with the initial condition $\langle \sigma (0)\rangle  = \langle \dot{\sigma}(0) \rangle  =0$, 
inferred from the QCD preheating, and $n_N(0) =\bar{n}_N(0)=0$. 
$\langle \sigma \rangle $ arrives at the true vacuum, $\langle \sigma \rangle  = f_\pi = 92.4$ MeV, at $M_\sigma t \sim 8$. 
The nonadiabatic production of $(n_N + \bar{n}_N)$ ends 
when $M_\sigma t \sim 5$.} 
\label{sigma-t:fig}
\end{figure}

The backreaction terms correspond to the effective mass of $\sigma$ arising as the plasma effect through the nonadiabatic production of the pion and the nucleons due to the dynamic $\sigma$. 
We have checked that their effect on the motion of $\langle\sigma\rangle $ is sufficiently small. The nonadiabatic particle production occurs around $\langle\sigma\rangle \sim 0$ at once, but the parametric resonance does not  since the later motion does not come back there due to the friction $\gamma$.
Thus the dynamic $\langle \sigma \rangle $ is well approximately described 
only by Eq.(\ref{EOM-sigma}) without interaction terms with pions 
and nucleons. 
The time evolution of $\langle \sigma \rangle $ is plotted in 
Fig.~\ref{sigma-t:fig}, as a function of $(M_\sigma t)$.

Since the dynamic $\langle \sigma \rangle $ instantaneously rolls down from  $\langle\sigma\rangle \sim 0$ to $f_\pi$ with the strong friction, 
the potential energy $V(\sigma=0)-V(\sigma=f_\pi)=\frac{1}{8}(M_\sigma^2+3m_\pi^2)f_\pi^2\simeq(135\:{\rm MeV})^4$ is converted into the radiation energy. 
This triggers reheating of the Universe and the reheating temperature 
can then be estimated as 
\textcolor{black}{
\begin{align}
 T_{\rm reh} 
 &= \left[\frac{\frac{\pi^2}{30}g_*(0.7 T_{\rm pc})^4+(135\:{\rm MeV})^4}{\frac{\pi^2}{30}g_*}\right]^{1/4} 
 \notag\\ 
& \simeq 119 \: {\rm MeV},
\end{align}
where we have 
assumed the relativistic degrees of freedom $g_*\simeq 14$ at  $T 
= 0.7 T_{\rm pc}\simeq 109\:{\rm MeV}$ when the dynamic $\langle \sigma \rangle $ starts to roll.  The produced thermal entropy density can also be estimated as
\begin{align} 
 \frac{2\pi^2}{45}g_*\cdot ( 119  \: {\rm MeV})^3 
  \simeq 1.24 \times 10^7 \: {\rm MeV}^3.
\label{s:estimate}
\end{align}
}

\section{Nonadiabatic production of nucleons due to the dynamic $\langle \sigma \rangle $} 

The dynamic $\langle \sigma \rangle $ controls the mass of the nucleons through the Yukawa interaction in Eq.(\ref{LSM-Lagrangian:eq}): 
\begin{equation}
 \frac{2m_N}{f_\pi}\bar{N}_LMN_R = \frac{m_N}{f_\pi}\langle\sigma\rangle \cdot \bar{N}_LN_R+\cdots.
\label{mN:sigma}
\end{equation} 
Thus the nucleon mass $\tilde{m}_N(t)=m_N\frac{\langle\sigma(t)\rangle }{f_\pi}$ varies in time, following the time evolution of $\langle\sigma\rangle $ depicted in Fig.~\ref{sigma-t:fig}. Such a time-varying mass causes the nonperturbative nucleon production when the adiabaticity is violated: $|\dot{\tilde{m}}_N/\tilde{m}_N^2|\gtrsim 1$.  This inequality leads to the production range in terms of the $\sigma$ motion as 
\begin{equation}
 |\sigma| \lesssim \sqrt{\frac{f_\pi \langle\dot{\sigma}\rangle }{m_N}}\simeq 42 \:{\rm MeV}, \label{production_area:eq}
\end{equation}
where we read $\langle \dot{\sigma}\rangle \simeq \sqrt{V(\sigma=0)-V(\sigma=f_\pi)}\sim (135\:{\rm MeV})^2$.
Hence the nonperturbative nucleon production would be completed within the range in Eq.~(\ref{production_area:eq}). In terms of Fig. \ref{sigma-t:fig}, 
it corresponds to
\begin{equation}
 M_\sigma t\lesssim 5 
 \label{non-adiabatic_region:eq}
\,. 
\end{equation}
The actual production time can be earlier than the number in Eq.(\ref{non-adiabatic_region:eq}) because the estimated velocity $\langle\dot{\sigma}\rangle $ would be smaller due to the friction $\gamma$.

The total nucleon number density can be evaluated as~\footnote{ 
$\vec{A}$ denotes the spatial component of the vector variable $A$.} 
\begin{eqnarray}
 n_N(t) + \bar{n}_N(t)
  &=& \sum_{\vec{k}}\frac{\langle \tilde{\rho}_N(\vec{k},t)\rangle }{\omega(k,t)}
  \,.  \label{total_baryon} 
\end{eqnarray}
for $N=p,n$.
Here $\tilde{\rho}_N(\vec{k},t)$ is the kinetic energy density
of the nucleon in momentum space, which is derived from the Hamiltonian as
\begin{eqnarray}
 \tilde{\rho}_N(\vec{k},t)
  &=& \frac{1}{V}\left[\bar{N}_L(\vec{k},L)\left(\vec{\gamma}\cdot \vec{k}+\tilde{m}_N(t)\right)N_L(\vec{k},t) \right. \nonumber \\
  & & \quad \left. + \bar{N}_R(\vec{k},t)\left(\vec{\gamma}\cdot \vec{k}+\tilde{m}_N(t)\right)N_R(\vec{k},t)\right] \nonumber \\
  & & + 2\omega(\vec{k},t)\,,
  \label{energy_density}
\end{eqnarray}
with $\omega(k,t)=\sqrt{|\vec{k}|^2+\tilde{m}_N^2(t)}$ being the  one-particle energy of the nucleon, 
$V=\int d^3x$ the space volume of the system, 
and $N_L(\vec{k},t)$ and  $N_R(\vec{k},t)$ the Fourier transformed Dirac fields. The last term in  Eq.(\ref{energy_density}) corresponds to the subtraction of the negative vacuum energy ($4\times \frac{1}{2}\omega(\vec{k},t)$).
The time evolution of the total baryon number density 
in Eq.(\ref{total_baryon}) is thus determined by solving coupled equations of 
motion for $N_L(\vec{k},t)$, $N_R(\vec{k},t)$, and $\langle \sigma(t) \rangle $, which is plotted in Fig.~\ref{sigma-t:fig}. 
For more details, see Appendix~\ref{details}.

We find that the total number density $(n_N + \bar{n}_N)$ is explosively generated by the nonadiabatic-time varying $\langle \sigma (t) \rangle $
and gets asymptotically saturated to be $ \simeq 10^5 \ {\rm MeV}^3$
in the time range $M_\sigma t\lesssim 5$. 
\textcolor{black}{
Note that this number density 
is much larger than the thermal equilibrium density at the reheating temperature $T_{\rm reh}\simeq  119$ MeV, 
$[(n_N + \bar{n}_N)]_{\rm EQ}\sim  10^3 \ {\rm MeV}^3$, 
hence becomes overabundant}. 
Actually, the overproduced nucleons can survive long enough during the whole reheating process: 
the relaxation time scale, at which the overproduced nucleons pair-annihilate,  
can be estimated as $M_\sigma \Delta t_{\rm relax}=M_\sigma(n_N \langle \sigma v \rangle )^{-1}\sim 700 $, 
or equivalently,  
\begin{align} 
\Delta t_{\rm relax} \sim 5 \times 10^{-7} \ {\rm fs} 
\, , \label{relax:time}
\end{align} 
where the static nucleon-pair annihilation 
cross section has been evaluated as a classical disc 
$\langle \sigma v \rangle  \sim 4\pi/m_N^2$ 
with the impact parameter $(1/m_N)$.  
Thus the QCD preheating is operative to stock a large number of nucleon and antinucleon pairs in out-of-equilibrium until the relaxation time. 
This is a novel picture of the thermal history 
in the QCD phase transition epoch.

This fact tempts us to consider the application to the baryogenesis, 
in which the Sakharov criteria are required: 
the baryon number violating interaction, C- and CP-violating interactions, and out-of-equilibrium condition~\cite{Sakharov:1967dj}.

Once the baryonic asymmetry $\epsilon \equiv (n_N - \bar{n}_N)/(n_N + \bar{n}_N)$ is provided by the other mechanism until the relaxation time, 
the QCD preheating can successfully generate the desired 
amount of the net baryon number of the Universe, 
\begin{align} 
Y_B \equiv \frac{(n_N + \bar{n}_N)}{s} \cdot \epsilon 
\sim 
10^{-10} \times \left( \frac{\epsilon}{10^{-8}}\right) 
\,, \label{YB}
\end{align}
where the entropy density in Eq.(\ref{s:estimate}) has been taken into account.

Here a salient feature is seen: the observed baryon 
number can be realized by a relatively smaller asymmetry 
than that accumulated in the thermal equilibrium 
due to the overproduced nucleons, which will not be washed out as long as the asymmetry generation completes before the system comes back to the thermal equilibrium.

It is true that QCD in the Standard Model cannot solely generate the asymmetry $\epsilon$ in Eq.(\ref{YB}) 
because of the absence of sufficient CP- and baryon number-violating interactions,  
but Eq.(\ref{YB}) opens a new roadmap as 
new baryogenesis involving a number of the new sector candidates coupled to the Standard Model: 
once an external sector transferring the CP and baryon number violations to QCD is hypothesized, the QCD preheating triggered by the dynamic $\langle \bar{q}q \rangle $ can 
create the net baryon asymmetry through Eq.(\ref{YB}), if and only if it ends until the relaxation time $M_\sigma \Delta t_{\rm relax} \sim 700$.

%


\section{Production of baryon asymmetry} 

As a benchmark, we introduce a class of models  
in which the baryogenesis makes the most of the QCD preheating to supply $\epsilon$ in Eq.(\ref{YB}).

\begin{figure}[t]
    \centering
    \includegraphics[scale=0.42]{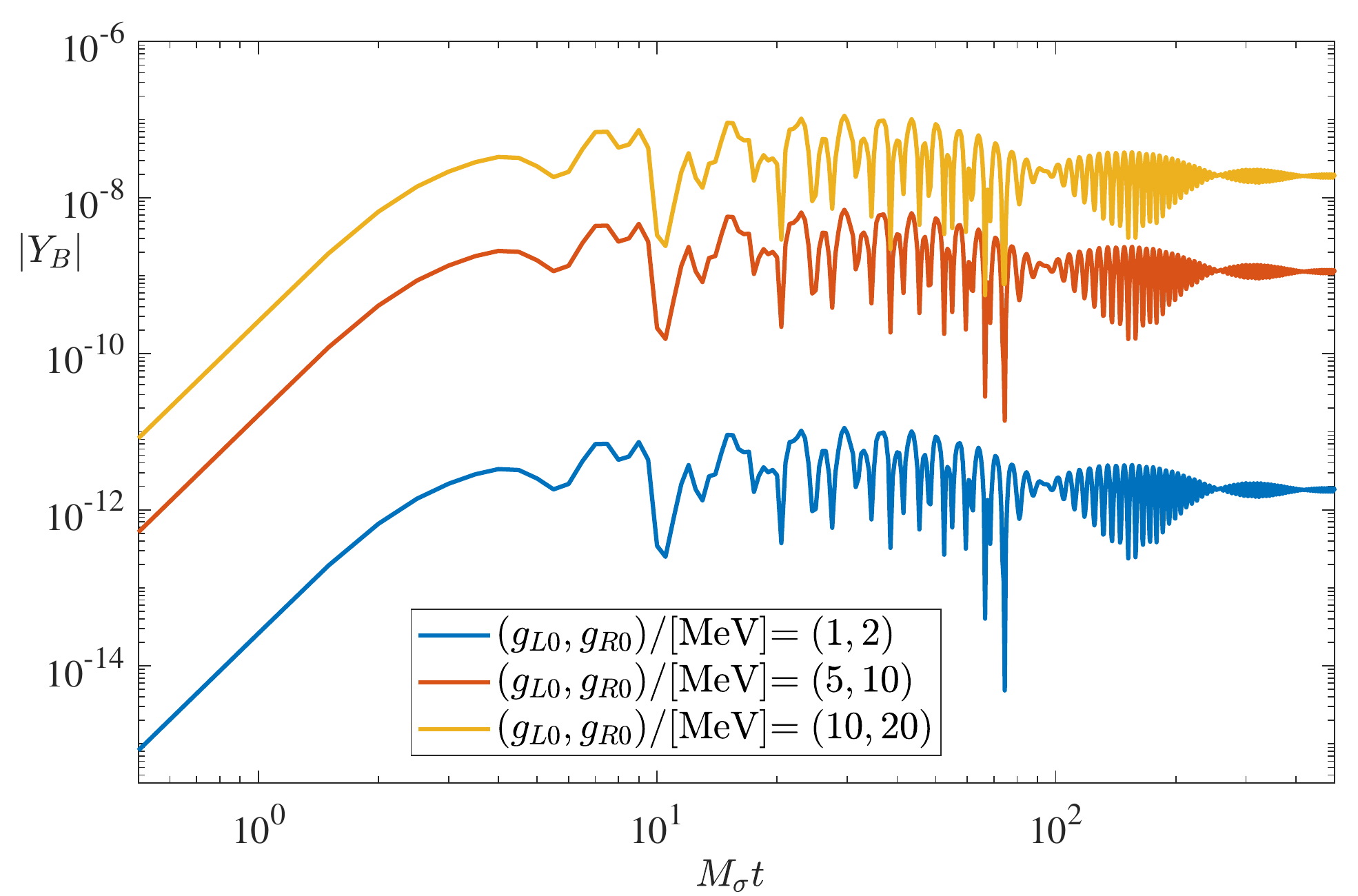}
    \caption{The time evolution of the baryon asymmetry yield $Y_B = n_B/s$, normalized with the entropy density $s= 10^7\,{\rm MeV}^3$. The model parameters have been fixed as described in the text.  
    }
    \label{eta-t:fig}
\end{figure}

Consider a dark sector with dark-Majorana neutron fields $n_{D_{L,R}}$ being 
allowed to couple to the QCD neutron fields $n_{L,R}$ in a minimal way~\footnote{
\textcolor{black}{
The mixing angle between the dark sector and QCD via the $g_{L,R}$ couplings 
are of ${\cal O}(10^{-4})$ for benchmark models in Fig.~\ref{eta-t:fig}, 
so that the dark sector contributions to 
the successful QCD hadron physics and chiral phase transition are safely 
negligible}. 
}: 
\begin{align}
 & {\cal L}_{n\mathchar` n_D} 
 \notag\\ 
  &= 
    -m_D\bar{n}_{D_R}n_{D_L} 
  -\frac{1}{2} M_{L} \overline{n_{D_L}^c} n_{D_{L}}
 -\frac{1}{2} M_{R} \overline{n_{D_R}^c} n_{D_{R}}
  \notag \\
  & \quad 
  -g_{L} \bar{n}_{R} n_{D_{L}}-g_{R} \bar{n}_{D_R} n_L+ \text{ h.c.}
\,, \label{dark:Lag}
\end{align}
where the superscript $c$ stands for the charge conjugation. 
In general, only two CP-violating phases can be physical to be introduced among the mass couplings ($m_D, g_L, g_R, M_L, M_R$).

In the dark sector scenario described in Eq.(\ref{dark:Lag}), the dynamic $\langle\sigma (t) \rangle $ also plays an important role in generating the CP-violating source.
Indeed, even the physical CP phases on the mass parameters can be erased by the appropriate diagonalization of the mass matrix since the Lagrangian consists of only the two-point interactions.  However, the neutron mass 
currently depends on time via $\langle \sigma(t)\rangle $ as in Eq.(\ref{mN:sigma}), and thus the CP phases can re-appear from the kinetic terms as the so-called Berry connection.  
A similar discussion can be seen in \cite{Enomoto:2018yeu}.
The combined effect of the CP phase and the baryon number violating couplings thus enables the simultaneous asymmetric production of the neutron and antineutron by the dynamic $\langle \sigma \rangle $.

For simplicity, we turn off one of two phases and embed it in $M_L$. With this phase, the nonzero Majorana mass couplings thus transfer the CP violation and the 
baryon number violation into the QCD sector.   
The dark baryons need to be as heavy as the neutron, in such a way to make the dark-sector communication with the neutron operative 
in the particle production process.

A similar idea to transfer the baryon number violation from a dark sector into the QCD sector via introducing a neutron-dark neutron coupling has been discussed in the literature~\cite{Bringmann:2018sbs}, 
where the production mechanism of the number densities differs from the nonperturbative one, however.

We shall evaluate the net baryon number instead of the asymmetry $\epsilon$, 
which can be defined by the (approximately conserved) $U(1)$ Noether charge as~\footnote{
In the present dynamic system, it is actually less costly to work on the evaluation of the dynamical net nucleon number instead of a direct estimate of $\epsilon$. } 
\begin{align}
n_{B}(t)
&= \frac{1}{V} \int d^{3} x 
\left(\langle 
n_L^\dag n_L\rangle  + 
\langle 
n_R^\dag n_R\rangle 
-\sum_{\vec{k}}2\right)
\,, \label{net:def}
\end{align}
where 
the last term corresponds to the subtraction of the divergent part induced by the zero-point energy ($4\times \frac{1}{2}$). 
The time evolution of the vacuum expectation values 
in Eq.(\ref{net:def})
is evaluated by solving chained equations of 
motion for $n_{L,R}$ coupled to $n_{D_{L,R}}$ 
through Eq.(\ref{dark:Lag}), together with the dynamic $\langle \sigma (t) \rangle $ obeying Eq.(\ref{EOM-sigma}).

The definition of the net baryon number in Eq.(\ref{net:def}) should potentially include the proton contribution as well, which, however, we can safely ignore in our analysis. This is because the proton does not have the CP- and baryon number-violating interactions, and thus the net baryon number by the proton cannot be generated.

The sigma motion makes the net baryon number produced through the baryon-number violating couplings $g_{L,R}$.  
We have observed that the net number starts to oscillate even after the baryon number production. 
This is because of the presence of the $n$-$n_D$ and $n$-$\bar{n}$ oscillations that last eternally.  
To obtain the static net baryon number, 
the couplings $g_{L,R}$ connecting the neutrons and the dark sector need to somehow get damping in time or vanishing 
at the later era,
such as 
$g_{L,R}(t) = \varphi(t) g_{L0,R 0}$ 
with $\varphi(t) = e^{- \Gamma_\varphi t} \cos m_\varphi t$. 
In that case, since the magnitude of the $g_{L,R}$ couplings asymptotically and promptly drops to zero, any phenomenological and astrophysical bounds can safely be satisfied when the observation is made.

The damping-oscillation factor $\varphi(t)$ could arise when one considers an underlying picture of the $g_{L,R}$ mass-mixing coupling to be given by a scalar or dilaton,  which communicates between the dark sector and the normal QCD sector. 
In that case, $\varphi(t)$ would be regarded as a background part of the scalar field, and the form of the damping oscillation in time would 
be provided by the scalar decay width $\Gamma_\varphi$, following 
the time evolution equation: 
$\langle\ddot{\varphi}\rangle  + \Gamma_\varphi \langle\dot{\varphi}\rangle  + m_\varphi^2 \langle\varphi\rangle  =0$. 
This ultraviolet completion could be rich in phenomenology and cosmology, which  is, however, beyond the current scope, to be pursued elsewhere.

Figure~\ref{eta-t:fig} shows the time evolution of 
the created net-baryon number yield $Y_B=n_B/s$, normalized by the entropy density $s  =10^7\,{\rm MeV}^3$ [see Eq.(\ref{s:estimate})]. 
The plot has been made  
for $M_R= 1000\, {\rm MeV}$, 
$M_L=1000 \cdot e^{i \frac{\pi}{3}}\, {\rm MeV}$, 
$m_D=1050$ MeV, 
$m_N=1000$ MeV,  
$m_\varphi = 10$ MeV, 
and $\Gamma_\varphi = 5$ MeV, 
along with various sets of the initial $g_{L,R}$ coupling strengths,   
$(g_{L0}, g_{R0})/{\rm MeV} = (1, 2), (5, 10)$, and $(10, 20)$. 
The detailed tools and methods for the present numerical computation 
are supplemented in Appendix~\ref{details}.

We have found that the net baryon number scales like $n_{B} \propto |g_L g_R|^2$. 
The net baryon number is essentially produced by the nonadiabatic $\sigma$ motion
and the associated nonperturbative nucleon-antinucleon pair production
for the time scale in the first phase 
$0 \le M_\sigma t \lesssim 5$, 
which happens before 
$\langle \sigma \rangle $ reaches the true vacuum at $M_\sigma t \sim 10$  
[See Fig.~\ref{sigma-t:fig}].  
This production-time range is consistent with a rough estimation done in Eq.(\ref{non-adiabatic_region:eq}). 
The time evolution of $Y_B$ turns to the second phase ($5 \lesssim M_\sigma t \lesssim 100$),  
where 
the intrinsic neutron-antineutron oscillation governs the dynamics. 
In the end, coming to the third phase 
($M_\sigma t \gtrsim 100$),  
$Y_B$ asymptotically gets saturated to the constant 
due to the damping oscillation of the 
couplings $g_{L,R}$, which start to operate at around $M_\sigma t \sim M_\sigma/\Gamma_\varphi= 100$. 
The general feature of $Y_B$ as viewed in Fig.~\ref{eta-t:fig} is qualitatively independent of $m_\varphi$ 
when it is lower compared with the particle production time scale $M_\sigma/5$ 
because $g_L$ and $g_R$ can maintain almost constants during the net baryon number production.  
On the other hand, the higher mass, $m_\varphi \gg M_\sigma /5$, might spoil the net number production 
because $g_L$ and $g_R$ become ineffective due to their violent vibrations around zero.

The asymmetry production is thus completed 
faster than the aforementioned relaxation time ($M_\sigma \Delta t_{\rm relax} \sim 700$). 
Thereby the baryogenesis is successfully accomplished 
to yield the desired amount of the baryon asymmetry, $Y_B = n_B/s 
\sim 10^{-10} - 10^{-9}$.  
We would remark that 
the typical baryogenesis scenario requires a high-scale physics, 
such as much higher than the electroweak scale, 
whereas the present one is realized at the lower scale due to the QCD preheating.

\section{Conclusions}  

The QCD preheating 
provides a nonperturbative cosmological particle production out of equilibrium and reheats the Universe 
(Fig.~\ref{rough_sketch}). 
Consequently,
the baryon asymmetry of 
the Universe can successfully be yielded through Eq.(\ref{YB}),  
once a source of the asymmetry is provided
before the system comes back to the thermal equilibrium, which corresponds to  the time scale $\Delta t_{\rm relax} \sim 10^{-7} \ {\rm fs}$ after 
the QCD preheating is completed.

The QCD preheating makes it possible to create 
not only the ordinary matters, 
but also even the net baryon number or non-standard particles if there is the new physics communicated with QCD.

The presently proposed QCD-scale baryogenesis can satisfy one of the most difficult Sakharov criteria, 
namely out-of-equilibrium condition, 
without introducing new physics effects. 
  This is in sharp contrast to the conventional way  invoking the first-order electroweak phase transition or 
heavy particle decays. 
Even the baryon number violation can be achieved around the QCD scale, not higher scales. 
Thus the QCD preheating does substitute the conventional methodologies to create 
the cosmological matter abundance observed in the present-day Universe, including the baryon asymmetry of the Universe, 
hence 
is a new paradigm for baryogenesis and paves the way for a new frontier 
involving variant Beyond the Standard Model candidates.

Another QCD baryogenesis has been addressed based on higher scale features of QCD, not at the subatomic scale~\cite{Ipek:2018lhm,Servant:2014bla,Croon:2019ugf}, 
where the production mechanism of the number densities is neither due to the nonperturbative time-dependent varying vacuum.

We introduced a class of the benchmark scenario to incorporate sufficient C- and CP-violating and baryon number violating interactions just as a reference, 
in which the generation of the baryonic asymmetry makes the most of the QCD preheating, 
and showed that the baryogenesis works (Fig.~\ref{eta-t:fig}).   
\textcolor{black}{
Full embedding of the dark and the Standard Model sectors into the scalegenesis 
would be worth pursuing, which is to be explored elsewhere. 
}

The QCD preheating leads to a sort of extra reheating in the thermal history, as illustrated in Fig.~\ref{rough_sketch}, hence has an impact also 
on exploring thermal or cosmological scenarios addressing 
the epoch around the QCD phase transition. 
It would be, for instance, worth revisiting the estimate of the cosmological abundance of the QCD axion, along with the QCD preheating.

\textcolor{black} 
{
The QCD preheating follows presence of the QCD supercooling as described in Sec.II, which would be subsequent supercooling involving the Standard Model Higgs and dark sectors with a scale invariant setup. 
Precise realization of the QCD supercooling thus requires nonperturbative analysis on the vacuum structure in massless-six flavor QCD and its tunneling rate. This issue deserves to another publication. 
}


\section*{Acknowledgements} 
We thank Mamiya Kawaguchi for useful comment on the QCD phase transition. 
This work was supported in part by the National Science Foundation of China (NSFC) under Grant No.11975108, 12047569, 12147217 
and the Seeds Funding of Jilin University (S.M.).  

\appendix

\section{Preliminaries for numerical evaluation of time evolution of the baryon number} 
\label{details}

In this Appendix we give a list of tools for the analysis on 
the QCD preheating on the basis of the linear sigma model description and 
a dark sector introduced in the main text.
Note that we formulate the fermion sector by the two-component spinors.
The relations to the Dirac fermions are followings;
\begin{equation}
 n=\left(\begin{array}{c}n_L \\ n_R^{c\dagger} \end{array} \right), \qquad n_D=\left(\begin{array}{c}n_{D_L} \\ n_{D_R}^{c\dagger} \end{array} \right),
\end{equation}
where the superscript ``$c$'' denotes the charge conjugate.

\begin{widetext} 

\subsection{Equations of motion}

The relevant equations of motion are: 
\begin{equation}
\begin{aligned}
0 &=\bar{\sigma}^{\mu} i \partial_{\mu} n_{L}-\tilde{m}_{N}^{*} n_{R}^{c \dagger}-g_{R} n_{D_{R}}^{c \dagger}\,,  \\
0 &=\sigma^{\mu} i \partial_{\mu} n_{L}^{\dagger} -\tilde{m}_{N} n_{R}^{c}-g_{R}^{*} n_{D_{R}}^{c}\,,  \\
0 &=\bar{\sigma}^{\mu} i \partial_{\mu} n_{R}^{c}-\tilde{m}_{N}^{*} n_{L}^{\dagger}-g_{L}^{*} n_{D_{L}}^{\dagger}\,,  \\
0 &=\sigma^{\mu} i \partial_{\mu} n_{L}^{\dagger}-\tilde{m}_{N} n_{R}^{c}-g_{L} n_{D_{L}}\,,  \\
0 &=\bar{\sigma}^{\mu} i \partial_{\mu} n_{D_{L}}-m_{D}^{*} n_{D_{R}}^{c \dagger}-M_{L}^{*} n_{D_{L}}^{\dagger}-g_{L}^{*} n_{R}^{c \dagger}\,,  \\
0 &=\sigma^{\mu} i \partial_{\mu} n_{D_{L}}^{\dagger}-m_{D} n_{D_{R}}^{c}-M_{L} n_{D_{L}}-g_{L} n_{R}^{c}\,,  \\
0 &=\bar{\sigma}^{\mu} i \partial_{\mu} n_{D_{R}}^{c}-m_{D}^{*} n_{D_{L}}^{\dagger}-M_{R}^{*} n_{D_{R}}^{c \dagger}-g_{R} n_{L}^{\dagger}\,,  \\
0 &=\sigma^{\mu} i \partial_{\mu} n_{D_{R}}^{c \dagger}-m_{D} n_{D_{L}}-M_{R} n_{D_{R}}^{c}-g_{R}^{*} n_{L}\,, 
\end{aligned}
\end{equation}
where $\tilde{m}_N(t)=m_N\frac{\langle\sigma(t)\rangle }{f_\pi}$. 
Equivalently, the above equations can collectively be represented by
\begin{equation}
 \bar{\sigma}^\mu i\partial_\mu \left( \begin{array}{c}n_L \\ n_R^c \\ n_{D_L} \\ n_{D_R}^c \end{array} \right) = \mathcal{M}^* \left( \begin{array}{c}n_L^\dagger \\ n_R^{c\dagger} \\ n_{D_L}^\dagger \\ n_{D_R}^{c\dagger} \end{array} \right), \qquad
 \sigma^\mu i\partial_\mu \left( \begin{array}{c}n_L^\dagger \\ n_R^{c\dagger} \\ n_{D_L}^\dagger \\ n_{D_R}^{c\dagger} \end{array} \right) = \mathcal{M} \left( \begin{array}{c}n_L \\ n_R^c \\ n_{D_L} \\ n_{D_R}^c \end{array} \right),
\end{equation}
where
\begin{equation}
 \mathcal{M} = \left( \begin{array}{cccc} 0 & \tilde{m}_N & 0 & g_R^* \\ \tilde{m}_N & 0 & g_L & 0 \\ 0 & g_L & M_L & m_D \\ g_R^* & 0 & m_D & M_R \end{array} \right). \label{eq:mass_matrix}
\end{equation}
Those can be transformed into the momentum space by the Fourier transform: 
\begin{equation}
\left(\begin{array}{c}
n_{L}(t, \vec{x}) \\
n_{R}^{c}(t, \vec{x}) \\
n_{D_{L}}(t, \vec{x}) \\
n_{D_{R}}^{c}(t, \vec{x})
\end{array}\right)_{\alpha}=\int \frac{d^{3} k}{(2 \pi)^{3}} e^{i \vec{k} \cdot \vec{x}} \sum_{s=\pm}\left(e_{\vec{k}}^{s}\right)_{\alpha}\left(\begin{array}{c}
n_{L, \vec{k}}^{s}(t) \\
n_{R, \vec{k}}^{s}(t) \\
D_{L, \vec{k}}^{s}(t) \\
D_{R, \vec{k}}^{s}(t)
\end{array}\right)
\,,  
\end{equation}
where 
\begin{equation}
\left(e_{\vec{k}}^{s}\right)_{1}=\sqrt{\frac{1}{2}\left(1+\frac{s k^{3}}{|\vec{k}|}\right)}, \quad\left(e_{\vec{k}}^{s}\right)_{2}=s e^{i \theta_{\vec{k}}} \sqrt{\frac{1}{2}\left(1-\frac{s k^{3}}{|\vec{k}|}\right)}
\,,\end{equation}
and 
\begin{equation}
e^{i \theta_{\vec{k}}} \equiv \frac{k^{1}+i k^{2}}{\sqrt{\left(k^{1}\right)^{2}+\left(k^{2}\right)^{2}}}
\,.
\end{equation}
Then we get 
\begin{equation}
\begin{aligned}
\partial_{t} n_{L, \vec{k}}^{s} &=i s k n_{L, \vec{k}}^{s}+i s e^{-i \theta_{\vec{k}}}\left(\tilde{m}_{N}^{*} n_{R,-\vec{k}}^{s \dagger}+g_{R} D_{R,-\vec{k}}^{s \dagger}\right)\,,  \\
\partial_{t} n_{R, \vec{k}}^{s} &=i s k n_{R, \vec{k}}^{s}+i s e^{-i \theta_{\vec{k}}}\left(\tilde{m}_{N}^{*} n_{L,-\vec{k}}^{s \dagger}+g_{L}^{*} D_{L,-\vec{k}}^{s \dagger}\right)\,,  \\
\partial_{t} D_{L, \vec{k}}^{s} &=i s k D_{L, \vec{k}}^{s}+i s e^{-i \theta_{\vec{k}}}\left(m_{D}^{*} D_{R,-\vec{k}}^{s \dagger}+M_{L}^{*} D_{L,-\vec{k}}^{s \dagger}+g_{L}^{*} n_{R,-\vec{k}}^{s \dagger}\right)\,,  \\
\partial_{t} D_{R, \vec{k}}^{s} &=i s k D_{R, \vec{k}}^{s}+i s e^{-i \theta_{\vec{k}}}\left(m_{D}^{*} D_{L,-\vec{k}}^{s \dagger}+M_{R}^{*} D_{R,-\vec{k}}^{s \dagger}+g_{R} n_{L,-\vec{k}}^{s \dagger}\right)\,.
\end{aligned}
\label{EOMs}
\end{equation} 
Because of the spacial-homogeneous Universe we are allowed to focus only on one mode ($n_k$) having $\vec{k}=\left(\epsilon, 0, k_{z}\right)$ with $\epsilon \rightarrow 0$, which leads to $e^{i \theta_{\vec{k}}}=1$, and then perform the momentum-space integral $\int 4\pi  k^2 n_k dk$.

\subsection{Time evolution of two-point functions and the total/net baryon number}

The total number densities of neutron can be read off from Eqs.(\ref{total_baryon}) and (\ref{energy_density} as:  
\begin{equation}
n_n+\bar{n}_n= \frac{1}{V}\int\frac{d^3k}{(2\pi)^3}\sum_{s=\pm} \left(1-\frac{s|\vec{k}|}{\omega_k}\left(\langle n_{L,\vec{k}}^{s\dagger}n_{L,\vec{k}}^s\rangle+\langle n_{R,\vec{k}}^{s\dagger}n_{R,\vec{k}}^s\rangle\right)  +\frac{\tilde{m}_N}{\omega_k}\left(se^{i\theta_{\vec{k}}}\langle n_{R,-\vec{k}}^s n_{L,\vec{k}}^s\rangle+se^{-i\theta_{\vec{k}}}\langle n_{L,\vec{k}}^{s\dagger} n_{R,-\vec{k}}^{s\dagger}\rangle\right)\right)
\end{equation}
where $V=\int d^3x$ is a volume of the system and
\begin{equation}
\omega_k(t)=\sqrt{|\vec{k}|^2+\tilde{m}_N(t)^2}.
\end{equation}

The net baryon number density in Eq.(\ref{net:def}) 
is rewritten in terms of the Fourier transformed fields as 
\begin{equation}
\begin{aligned}
n_{B} &=\frac{1}{V} \int d^{3} x \frac{1}{2}\left(\left\langle n_{L}^{\dagger} \bar{\sigma}^{0} n_{L}\right\rangle -\left\langle n_{L} \sigma^{0} n_{L}^{\dagger}\right\rangle -\left\langle n_{R}^{c \dagger} \bar{\sigma}^{0} n_{R}^{c}\right\rangle +\left\langle n_{R}^{c} \sigma^{0} n_{R}^{c \dagger}\right\rangle \right) \\ &=\frac{1}{V} \int \frac{d^{3} k}{(2 \pi)^{3}} \frac{1}{2} \sum_{s=\pm}\left(\left\langle n_{L, \vec{k}}^{s \dagger} n_{L, \vec{k}}^{s}\right\rangle -\left\langle n_{L,-\vec{k}}^{s} n_{L,-\vec{k}}^{s \dagger}\right\rangle \right).\\ &=\frac{1}{V} \int \frac{d^{3} k}{(2 \pi)^{3}} \sum_{s=\pm}\left(\left\langle n_{L, \vec{k}}^{s \dagger} n_{L, \vec{k}}^{s}\right\rangle -\left\langle n_{R, \vec{k}}^{s \dagger} n_{R, \vec{k}}^{s}\right\rangle \right) 
\end{aligned}
\,, 
\end{equation} 
The two-point functions relevant to this net baryon number density 
are developed in time through the time evolution equations given in Eq.(\ref{EOMs}):  
\begin{align}
\pt \langle \nLkd \nLk \rangle  
&= 
-is \eiqk 
\left( 
\tilmN \langle \nRmk \nLk \rangle  
+ 
g_R^\ast 
\langle \DRmk \nLk \rangle  
\right)
+ {\rm h.c.}\,,\\
\pt \langle \nRmk \nLk \rangle  
&= 
2 i s \abk \langle \nRmk \nLk \rangle  
- 
is \eiqmk \left( \tilmNs \langle \nLkd \nLk \rangle  + g_L^\ast \langle \DLkd \nLk \rangle  \right) \notag\\
&\hspace{5mm}+ 
is \eiqmk \left( \tilmNs \langle \nRmk \nRmkd \rangle  + g_R \langle \nRmk \DRmkd \rangle  \right)\,,\\
\pt \langle \DRmk \nLk \rangle  
&= 
2is \abk \langle \DRmk \nLk \rangle  
+ 
is \eiqmk \left( \tilmNs \langle \DRmk \nRmkd \rangle  + g_R\langle \DRmk \DRmkd \rangle  \right) \notag\\
&\hspace{5mm}- 
is \eiqmk \left( \mDs \langle \DLkd \nLk \rangle  + \MRs \langle \DRkd \nLk \rangle  + g_R \langle \nLkd \nLk \rangle  \right)\,,\\
\pt \langle \DLkd \nLk \rangle  
&= 
is \eiqmk \left( \tilmNs \langle \DLkd \nRmkd \rangle  + g_R \langle \DLkd \DRmkd \rangle  \right) \notag\\
&\hspace{5mm}- 
is \eiqk \left( \mD \langle \DRmk \nLk \rangle  + \ML \langle \DLmk \nLk \rangle  + g_L \langle \nRmk \nLk \rangle  \right)\,,\\
\pt \langle \nRmkd \nRmk \rangle 
&= 
is \eiqk \left( \tilmN \langle \nLk \nRmk \rangle  + g_L \langle \DLk \nRmk \rangle  \right) 
+ {\rm h.c.}\,,\\
\pt \langle \DRmkd \nRmk \rangle  
&= 
is \eiqk \left( \mD \langle \DLk \nRmk \rangle  + \MR \langle \DRk \nRmk \rangle  + g_R \langle \nLk \nRmk \rangle  \right) \notag\\
&\hspace{5mm}- 
is \eiqmk \left( \tilmNs \langle \DRmkd \nLkd \rangle  + g_L^\ast \langle \DRmkd \DLkd \rangle  \right)\,,\\
\pt \langle \DRmkd \DRmk \rangle  
&= 
is \eiqk \left( \mD \langle \DLk \DRmk \rangle  + \MR \langle \DRk \DRmk \rangle  + g_R^\ast \langle \nLk \DRmk \rangle  \right) 
+ {\rm h.c.}\,,\\
\pt \langle \DRkd \nLk \rangle  
&= 
-is \eiqk \left( \mD \langle \DLmk \nLk \rangle  + \MR \langle \DRmk \nLk \rangle  + g_R^\ast \langle \nLmk \nLk \rangle  \right)\notag\\
&\hspace{5mm}+ 
is \eiqmk \left( \tilmNs \langle \DRkd \nRmkd \rangle  + g_R \langle \DRkd \DRmkd \rangle  \right)\,,\\
\pt \langle \DLk \nRmk \rangle  
&= 
2is \abk \langle \DLk \nRmk \rangle  +
is \eiqmk \left( \mDs \langle \DRmkd \nRmk \rangle  + \MLs \langle \DLmkd \nRmk \rangle  + g_L^\ast \langle \nRmkd \nRmk \rangle  \right)\notag\\
&\hspace{5mm}- 
is \eiqmk \left( \tilmNs \langle \DLk \nLkd \rangle  + g_L^\ast \langle \DLk \DLkd \rangle  \right)\,,\\
\pt \langle \DRk \nRmk \rangle  
&= 
2is \abk \langle \DRk \nRmk \rangle  -is \eiqmk \left( \tilmNs \langle \DRk \nLkd \rangle  + g_L^\ast \langle \DRk \DLkd \rangle  \right) \notag\\
&\hspace{5mm}+ 
is \eiqmk \left( \mDs \langle \DLmkd \nRmk \rangle  + \MRs \langle \DRmkd \nRmk \rangle  + g_R \langle \nLmkd \nRmk \rangle  \right)\,,\\
\pt \langle \DRmk \DLk \rangle  
&= 
2is \abk \langle \DRmk \DLk \rangle  -is \eiqmk \left( \mDs \langle \DLkd \DLk \rangle  + \MRs \langle \DRkd \DLk \rangle  + g_R \langle \nLkd \DLk \rangle  \right)\notag\\
&\hspace{5mm}+ 
is \eiqmk \left( \mDs \langle \DRmk \DRmkd \rangle  + \MLs \langle \DRmk \DLmkd \rangle  + g_L^* \langle \DRmk \nRmkd \rangle  \right)\,,\\
\pt \langle \DLmk \nLk \rangle  
&= 
2is \abk \langle \DLmk \nLk \rangle  + is \eiqmk \left( \tilmN \langle \DLmk \nRmkd \rangle  + g_R \langle \DLmk \DRmkd \rangle  \right)\notag\\
&\hspace{5mm}- 
is \eiqmk \left( \mDs \langle \DRkd \nLk \rangle  + \MLs \langle \DLkd \nLk \rangle  + g_L^* \langle \nRkd \nLk \rangle  \right)\,,\\
\pt \langle \nLmk \nLk \rangle  
&= 
2is \abk \langle \nLmk \nLk \rangle  - is \eiqmk \left( \tilmNs \langle \nRkd \nLk \rangle  + g_R \langle \DRkd \nLk \rangle  \right)\notag\\
&\hspace{5mm}+ 
is \eiqmk \left( \tilmNs \langle \nLmk \nRmkd \rangle  + g_R \langle \nLmk \DRmkd \rangle  \right)\,,\\
\pt \langle \DRk \DRmk \rangle  
&= 
2is \abk \langle \DRk \DRmk \rangle  - is \eiqmk \left( \mDs \langle \DRk \DLkd \rangle  + \MRs \langle \DRk \DRkd \rangle  + g_R \langle \DRk \nLkd \rangle  \right)\notag\\
&\hspace{5mm} +
is \eiqmk \left( \mDs \langle \DLmkd \DRmk \rangle  + \MRs \langle \DRmkd \DRmk \rangle  + g_R \langle \nLmkd \DRmk \rangle  \right)\,,\\
\pt \langle \DLmkd \nRmk \rangle  
&= 
is \eiqk \left( \mD \langle \DRk \nRmk \rangle  + \ML \langle \DLk \nRmk \rangle  + g_L \langle \nRk \nRmk \rangle  \right)\notag\\
&\hspace{5mm}- 
is \eiqmk \left( \tilmNs \langle \DLmkd \nLkd \rangle  + g_L^\ast \langle \DLmkd \DLkd \rangle  \right)\,,\\
\pt \langle \DLkd \DLk \rangle  
&= 
-is \eiqk \left( \mD \langle \DRmk \DLk \rangle  + \ML \langle \DLmk \DLk \rangle  + g_L \langle \nRmk \DLk \rangle  \right) 
+ {\rm h.c.}\,,\\
\pt \langle \nRk \nRmk \rangle  
&= 
2is \abk \langle \nRk \nRmk \rangle  + is \eiqmk \left( \tilmNs \langle \nLmkd \nRmk \rangle  + g_L^\ast \langle \DLmkd \nRmk \rangle  \right)\notag\\
&\hspace{5mm}- 
is \eiqmk \left( \tilmNs \langle \nRk \nLkd \rangle  + g_L^\ast \langle \nRk \DLkd \rangle  \right)\,,\\
\pt \langle \DLmk \DLk \rangle  
&= 
2is \abk \langle \DLmk \DLk \rangle  -is \eiqmk \left( \mDs \langle \DRkd \DLk \rangle  + \MLs \langle \DLkd \DLk \rangle  + g_L^\ast \langle \nRkd \DLk \rangle  \right)\notag\\
&\hspace{5mm}+ 
is \eiqmk \left( \mDs \langle \DLmk \DRmkd \rangle  + \MLs \langle \DLmk \DLmkd \rangle  + g_L^\ast \langle \DLmk \nRmkd \rangle  \right)\,,\\
\pt \langle \DLkd \DRk \rangle  
&= 
is \eiqmk\left( \mDs \langle \DLkd \DLmkd \rangle  + \MRs \langle \DLkd \DRmkd \rangle  + g_R \langle \DLkd \nLmkd \rangle  \right)\notag\\
&\hspace{5mm}- 
is \eiqk \left( \mD \langle \DRmk \DRk \rangle  + \ML \langle \DLmk \DRk \rangle  + g_L \langle \nRmk \DRk \rangle  \right)\,,\\
\pt \langle \nRkd \nLk \rangle  
&= 
-i s\eiqk \left( \tilmN \langle \nLmk \nLk \rangle  + g_L \langle \DLmk \nLk \rangle  \right)\notag\\
&\hspace{5mm}+ 
is \eiqmk \left( \tilmNs \langle \nRkd \nRmkd \rangle  + g_R \langle \nRkd \DRmkd \rangle  \right)\,. 
\end{align}

\subsection{Initial values}
To solve the sets of equations derived as above, we also need to place the initial conditions. 
We expect that the mass-eigenstate fields in the system initially behave as free fields since the fields do not have any other interactions except those due to $\langle \sigma \rangle $.  
It is convenient to introduce a collective Majonara field as 
\begin{equation}
\left[\psi_{\vec{k}}^{s }\right]^{i}=   
\left(\begin{array}{c}
n_{L, \vec{k}}^{s}(t) \\
n_{R, \vec{k}}^{s}(t) \\
D_{L, \vec{k}}^{s}(t) \\
D_{R, \vec{k}}^{s}(t)
\end{array}\right)^i
\,.
\end{equation} 
By defining 
\begin{equation}
\begin{aligned}
&{\left[A_{\vec{k}}\right]^{i j} \equiv \frac{1}{2 V} \sum_{s}\left(\left\langle\left[\psi_{\vec{k}}^{s \dagger}\right]^{i}\left[\psi_{\vec{k}}^{s}\right]^{j}\right\rangle -\left\langle\left[\psi_{-\vec{k}}^{s}\right]^{j}\left[\psi_{-\vec{k}}^{s \dagger}\right]^{i}\right\rangle \right)} \\
&{\left[\bar{A}_{\vec{k}}\right]^{i j} \equiv \frac{1}{2 V} \sum_{s} s\left(\left\langle\left[\psi_{\vec{k}}^{s \dagger}\right]^{i}\left[\psi_{\vec{k}}^{s}\right]^{j}\right\rangle -\left\langle\left[\psi_{-\vec{k}}^{s}\right]^{j}\left[\psi_{-\vec{k}}^{s \dagger}\right]^{i}\right\rangle \right)} \\
&{\left[B_{\vec{k}}\right]^{i j} \equiv \frac{1}{2 V} \sum_{s}\left(\left\langle s e^{i \theta_{\vec{k}}}\left[\psi_{-\vec{k}}^{s}\right]^{i}\left[\psi_{\vec{k}}^{s}\right]^{j}\right\rangle +\left\langle s e^{i \theta_{\vec{k}}}\left[\psi_{-\vec{k}}^{s}\right]^{j}\left[\psi_{\vec{k}}^{s}\right]^{i}\right\rangle \right)} \\
&{\left[\bar{B}_{\vec{k}}\right]^{i j} \equiv \frac{1}{2 V} \sum_{s} s\left(\left\langle s e^{i \theta_{\vec{k}}}\left[\psi_{-\vec{k}}^{s}\right]^{i}\left[\psi_{\vec{k}}^{s}\right]^{j}\right\rangle +\left\langle s e^{i \theta_{\vec{k}}}\left[\psi_{-\vec{k}}^{s}\right]^{j}\left[\psi_{\vec{k}}^{s}\right]^{i}\right\rangle \right)}
\end{aligned}
\,,  
\end{equation}  
their initial conditions at $t=t_0$ are then set as 
\begin{equation}
\begin{gathered}
{\left[A_{\vec{k}}\left(t_{0}\right)\right]^{i j}=\left[\bar{B}_{\vec{k}}\left(t_{0}\right)\right]^{i j}=0} \,, \\
{\left[\bar{A}_{\vec{k}}\left(t_{0}\right)\right]^{i j}=\sum_{\ell}\left[U^{*}\right]^{i \ell} \frac{|\vec{k}|}{\omega_{k}^{\ell}}\left[U^{T}\right]^{\ell j}} \,, \\
{\left[B_{\vec{k}}\left(t_{0}\right)\right]^{i j}=\sum_{\ell} [U]^{i \ell} \frac{\mathcal{M}_{d}^{\ell \ell}}{\omega_{k}^{\ell}}\left[U^{T}\right]^{\ell j}}
\,, 
\end{gathered}
\label{initial:condi}
\end{equation}
    where $\mathcal{M}_{d}$ is a diagonalized mass matrix related to  the original mass matrix $\mathcal{M}$
    in Eq.(\ref{eq:mass_matrix})
    through a unitary transformation by $U$ as 
\begin{equation}
\mathcal{M}_{d} \equiv U^{T} \mathcal{M} U 
\,. 
\end{equation}
In Eq.(\ref{initial:condi}) $\omega_{k}^{\ell}$ is the energy with the diagonalized mass as
\begin{equation}
    \omega_{k}^{\ell} \equiv \sqrt{|\mathrm{k}|^{2}+\left(\mathcal{M}_{d}^{\ell \ell}\right)^{2}} .
\end{equation}

In terms of the two-point functions, the initial conditions (set at $t_0=0$) read 
\begin{equation}
\begin{array}{lll}
\left\langle O_{1}\right\rangle_{t=0} & =\left\langle n_{L, \vec{k}}^{s \dagger} n_{L, \vec{k}}^{s}\right\rangle_{t=0} & =\frac{V}{2}\left(1+s\left[\bar{A}_{k}\left(t_{0}\right)\right]^{11}\right)\,,  \\
\left\langle O_{2}\right\rangle_{t=0} & =\left\langle n_{R,-\vec{k}}^{s} n_{L, \vec{k}}^{s}\right\rangle_{t=0} & =\frac{V s}{2}\left[B_{k}\left(t_{0}\right)\right]^{12}\,,  \\
\left\langle O_{3}\right\rangle_{t=0} & =\left\langle D_{R,-\vec{k}}^{s} n_{L, \vec{k}}^{s}\right\rangle_{t=0} & =\frac{V s}{2}\left[B_{k}\left(t_{0}\right)\right]^{41}\,,  \\
\left\langle O_{4}\right\rangle_{t=0} & =\left\langle D_{L, \vec{k}}^{s \dagger} n_{L, \vec{k}}^{s}\right\rangle_{t=0} & =\frac{V}{2}\left(s\left[\bar{A}_{k}\left(t_{0}\right)\right]^{31}\right)\,,  \\
\left\langle O_{5}\right\rangle_{t=0} & =\left\langle n_{R,-\vec{k}}^{s \dagger} n_{R,-\vec{k}}^{s}\right\rangle_{t=0} & =\frac{V}{2}\left(1+s\left[\bar{A}_{k}\left(t_{0}\right)\right]^{22}\right)\,,  \\
\left\langle O_{6}\right\rangle_{t=0} & =\left\langle D_{R,-\vec{k}}^{s \dagger} n_{R,-\vec{k}}^{s}\right\rangle_{t=0} & =\frac{V}{2}\left(s\left[\bar{A}_{\boldsymbol{k}}\left(t_{0}\right)\right]^{42}\right)\,,  \\
\left\langle O_{7}\right\rangle_{t=0} & =\left\langle D_{R,-\vec{k}}^{s+} D_{R,-\vec{k}}^{s}\right\rangle_{t=0} & =\frac{V}{2}\left(1+s\left[\bar{A}_{k}\left(t_{0}\right)\right]^{44}\right)\,,  \\
\left\langle O_{8}\right\rangle_{t=0} & =\left\langle D_{R, \vec{k}}^{s t} n_{L, \vec{k}}^{s}\right\rangle_{t=0} & =\frac{V}{2}\left(s\left[\bar{A}_{k}\left(t_{0}\right)\right]^{41}\right)\,,  \\
\left\langle O_{9}\right\rangle_{t=0} & =\left\langle D_{L, \vec{k}}^{s} n_{R,-\vec{k}}^{s}\right\rangle_{t=0} & =-\frac{V s}{2}\left[B_{k}\left(t_{0}\right)\right]^{32}\,,  \\
\left\langle O_{10}\right\rangle_{t=0} & =\left\langle D_{R, \vec{k}}^{s} n_{R,-\vec{k}}^{s}\right\rangle_{t=0} & =-\frac{V s}{2}\left[B_{k}\left(t_{0}\right)\right]^{42}\,,  \\
\left\langle O_{11}\right\rangle_{t=0} & =\left\langle D_{R,-\vec{k}}^{s} D_{L, \vec{k}}^{s}\right\rangle_{t=0} & =\frac{V s}{2}\left[B_{k}\left(t_{0}\right)\right]^{34}\,,  \\
\left\langle O_{12}\right\rangle_{t=0} & =\left\langle D_{L,-\vec{k}}^{s} n_{L, \vec{k}}^{s}\right\rangle_{t=0} & =\frac{V s}{2}\left[B_{k}\left(t_{0}\right)\right]^{31}\,,  \\
\left\langle O_{13}\right\rangle_{t=0} & =\left\langle n_{L,-\vec{k}}^{s} n_{L, \vec{k}}^{s}\right\rangle_{t=0} & =\frac{V s}{2}\left[B_{k}\left(t_{0}\right)\right]^{11}\,,  \\
\left\langle O_{14}\right\rangle_{t=0} & =\left\langle D_{R, \vec{k}}^{s} D_{R,-\vec{k}}^{s}\right\rangle_{t=0} & =-\frac{V s}{2}\left[B_{k}\left(t_{0}\right)\right]^{44}\,,  \\
\left\langle O_{15}\right\rangle_{t=0} & =\left\langle D_{L,-\vec{k}}^{s \dagger} n_{R,-\vec{k}}^{s}\right\rangle_{t=0} & =\frac{V}{2}\left(s\left[\bar{A}_{k}\left(t_{0}\right)\right]^{32}\right)\,,  \\
\left\langle O_{16}\right\rangle_{t=0} & =\left\langle D_{L, \vec{k}}^{s \dagger} D_{L, \vec{k}}^{s}\right\rangle_{t=0} & =\frac{V}{2}\left(1+s\left[\bar{A}_{k}\left(t_{0}\right)\right]^{33}\right)\,,  \\
\left\langle O_{17}\right\rangle_{t=0} & =\left\langle n_{R, \vec{k}}^{s} n_{R,-\vec{k}}^{s}\right\rangle_{t=0} & =-\frac{V s}{2}\left[B_{k}\left(t_{0}\right)\right]^{22}\,,  \\
\left\langle O_{18}\right\rangle_{t=0} & =\left\langle D_{L,-\vec{k}}^{s} D_{L, \vec{k}}^{s}\right\rangle_{t=0} & =\frac{V s}{2}\left[B_{k}\left(t_{0}\right)\right]^{33}\,,  \\
\left\langle O_{19}\right\rangle_{t=0} & =\left\langle D_{L, \vec{k}}^{s \dagger} D_{R, \vec{k}}^{s}\right\rangle_{t=0} & =\frac{V}{2}\left(s\left[\bar{A}_{k}\left(t_{0}\right)\right]^{34}\right)\,,  \\
\left\langle O_{20}\right\rangle_{t=0} & =\left\langle n_{R, \vec{k}}^{s \dagger} n_{L, \vec{k}}^{s}\right\rangle_{t=0} & =\frac{V}{2}\left(s\left[\bar{A}_{k}\left(t_{0}\right)\right]^{12}\right)^{*}\,.
\end{array}
\end{equation}

\end{widetext}


\end{document}